\documentclass[12pt,preprint]{emulateapj}
\usepackage[dvips]{color}

\newcommand{\msun}{$M_\sun$}

\newcommand{\zdrop}{$z_{850}-$dropout}
\newcommand{\ydrop}{$Y_{105}-$dropout}

\newcommand{\yy}{$Y_{105}$}
\newcommand{\jj}{$J_{125}$}
\newcommand{\hh}{$H_{160}$}

\shorttitle{Stellar masses and ages to $z=7-8$}
\shortauthors{I. Labb\'e et al.}

\def\figstack{
\begin{figure}
\centering
$$ $$ \includegraphics[width=7cm]{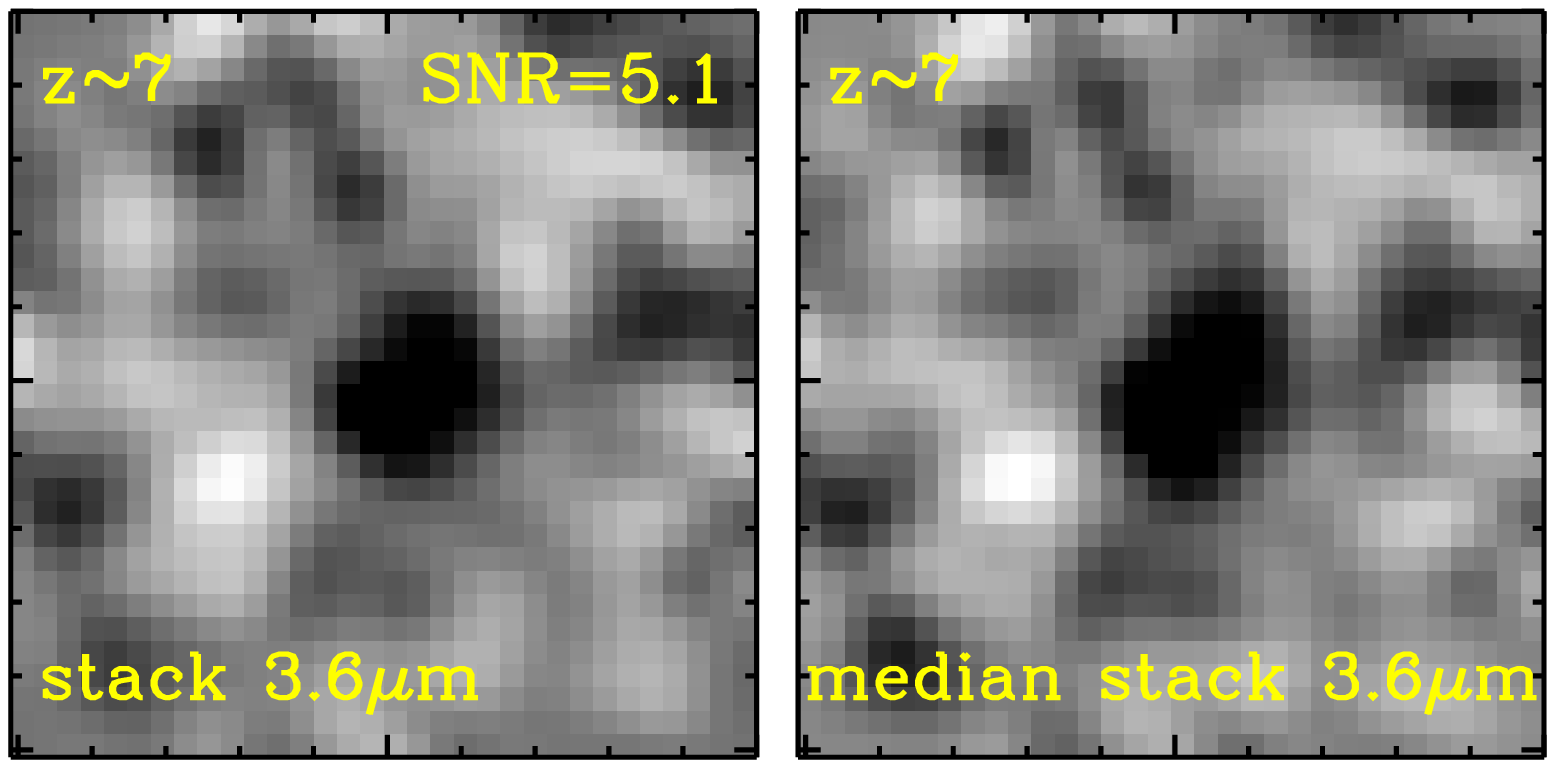}
\includegraphics[width=7cm]{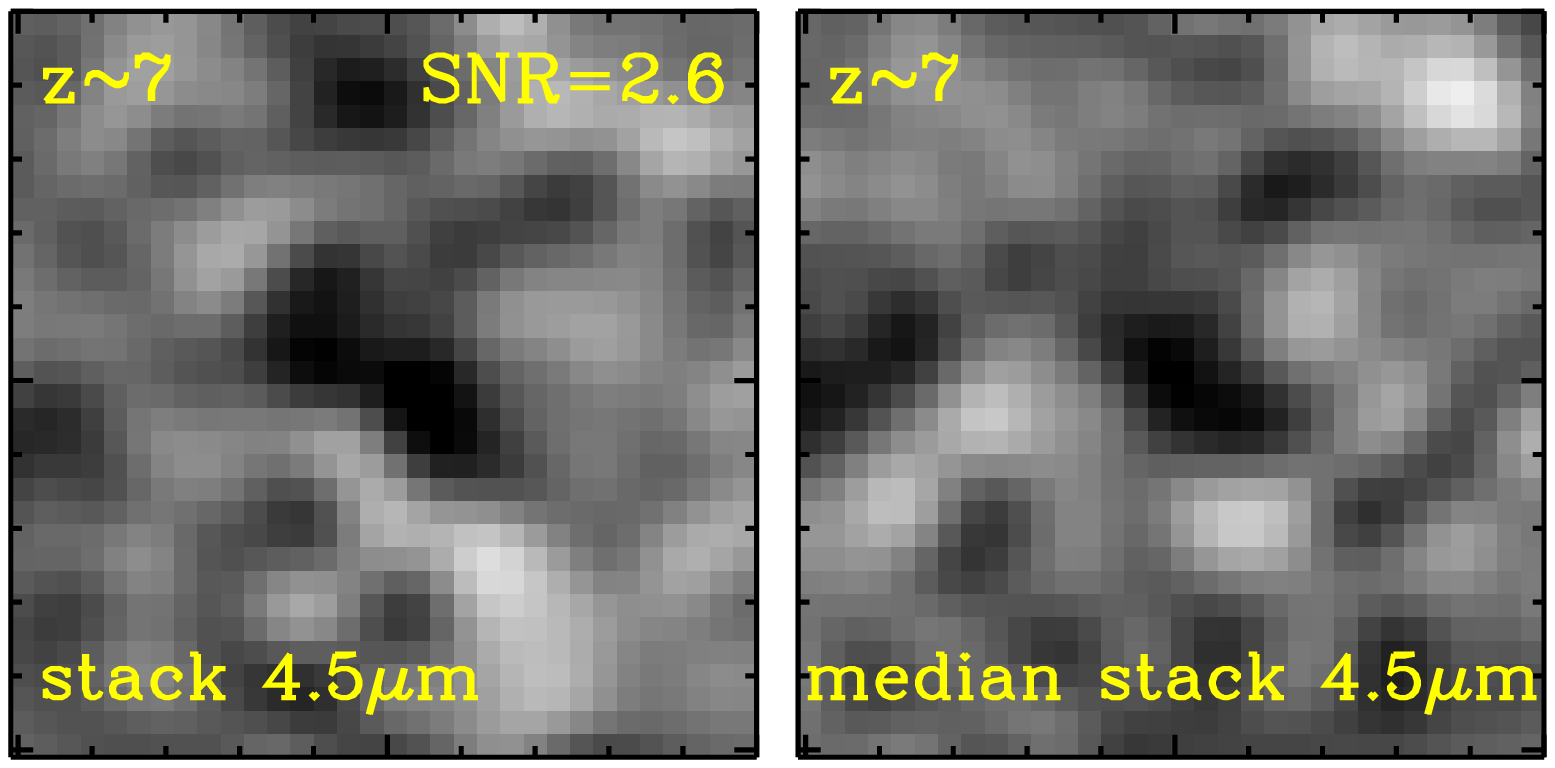}
\includegraphics[width=7cm]{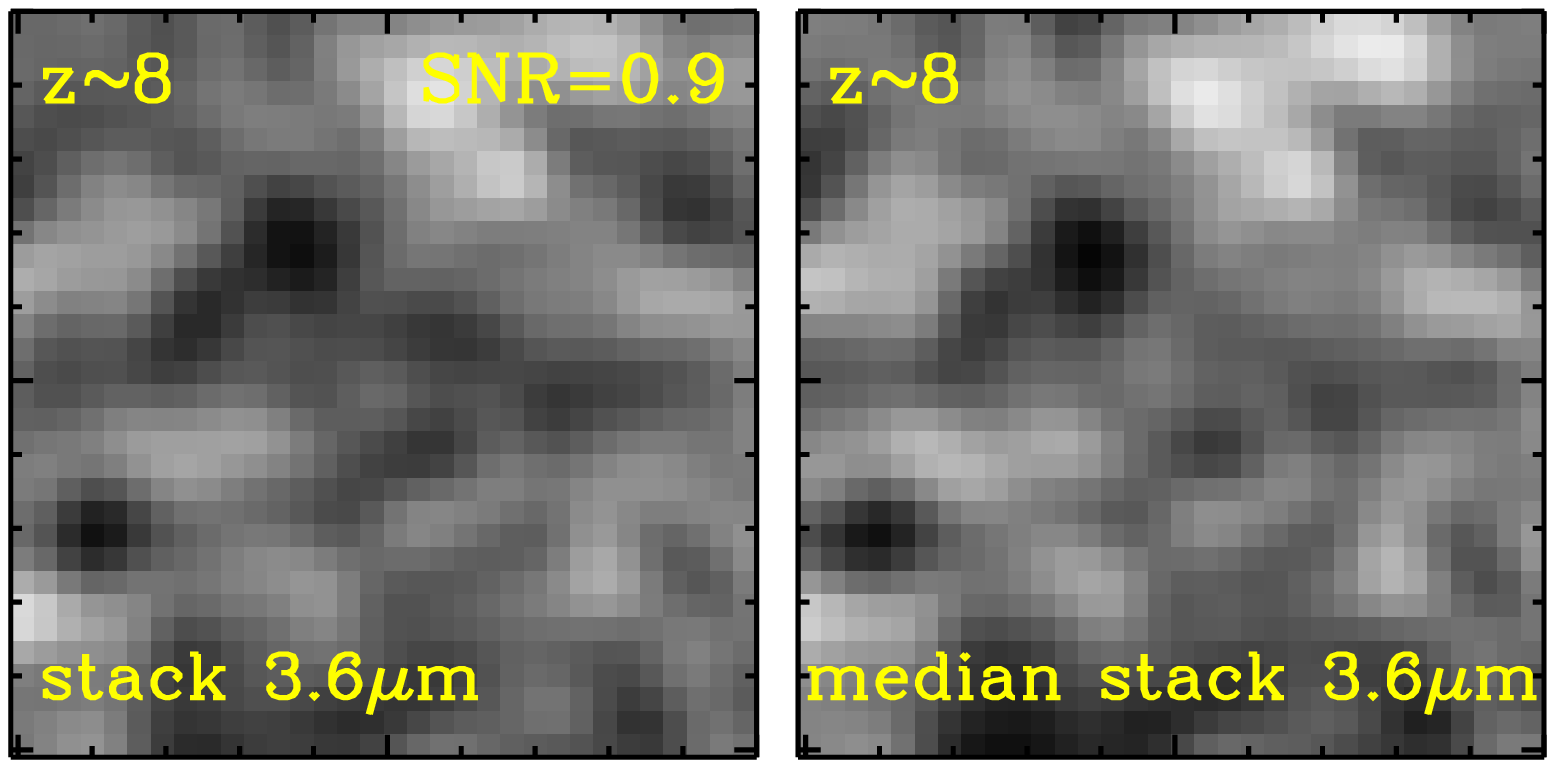} $$ $$
\caption{Stacked images of $z\sim7$ $z_{850}$-dropout and $z\sim8$ $Y_{105}$-dropout 
sources in the HUDF. The stacked $z_{850}$-dropouts in the $[3.6]-$band ($top\ left$) 
shows an unambiguous detection ($S/N=5.1$). The
median stack also shows a strong signal, indicating that it is not 
dominated by outliers. The S/N is calculated by bootstrap resampling. 
The $z_{850}$-dropouts stack is marginally detected ($S/N=2.6$) in $[4.5]$ ($middle\ row$).
The $Y_{105}$-dropouts stack ($bottom\ row$) shows no source in the center. 
Panels are in inverted grayscale and $9\farcs9\times9\farcs9$ on a side. 
\label{figstack}}
\end{figure}
}

\def\figsed{
\begin{figure*}
\epsscale{1.2}
\centering
\plotone{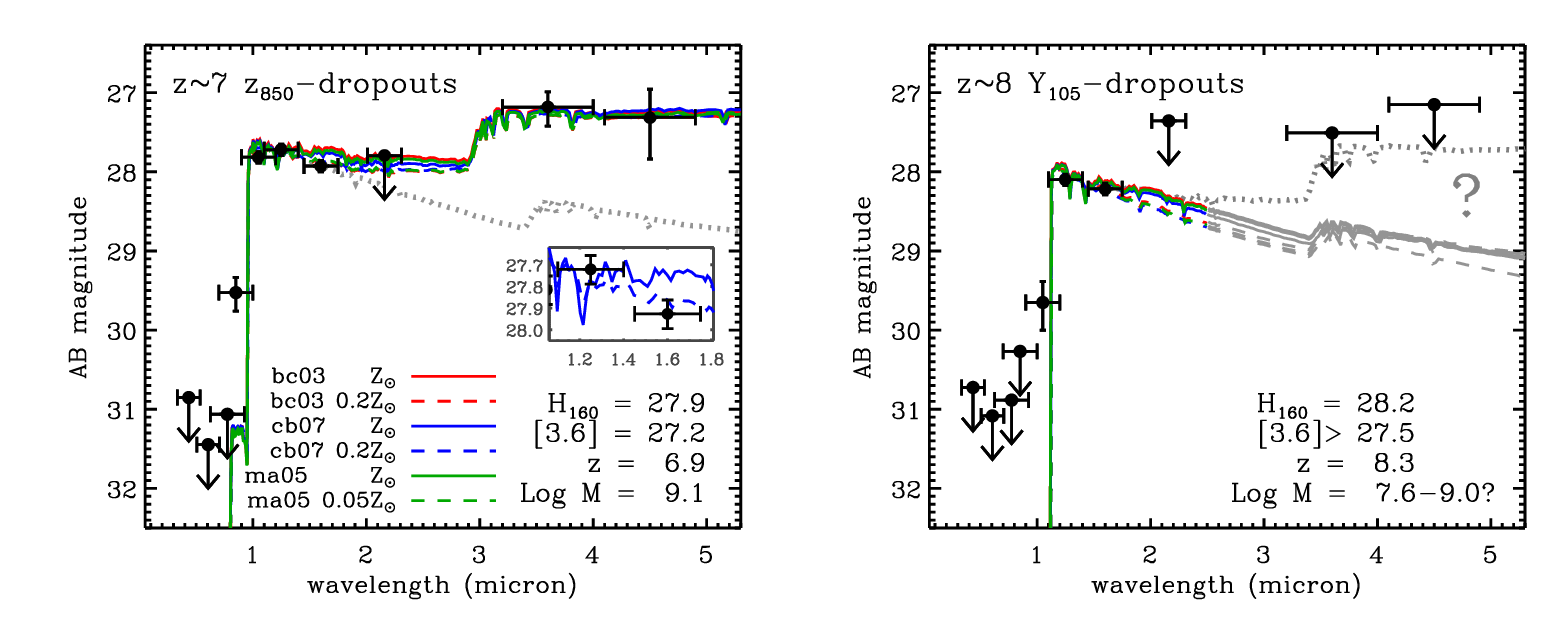}
\caption{Average broadband ACS, WFC3/IR, and IRAC SEDs of the 
$z\sim7$ and $z\sim8$  galaxies. The left panel shows the average 
$z\sim7$ $z_{850}$-dropout SED and the best-fit stellar population models. 
The blue far-UV slope and red $H_{160}-[3.6]$ color indicates a modest
Balmer break, expected for evolved stellar populations ($>100$Myr).
Excluding the IRAC measurements leads to $1.5$ dex smaller mass ($dotted\ line$).
These faint galaxies have similar $M/L$ and age as the luminous 
$z\sim7$ sources presented by \citet{Go09}. {The inset presents part of
the SED around $1.4\mu$m ($\sim1700\AA$ rest-frame) showing that the
best-fit sub-Solar metallicity models reproduce the quite blue far-UV 
slope better than do the best-fit Solar models
(see, e.g., \citealt{Bo09b} for more details).}
Differences between the best-fit parameters of 
various stellar population models {at fixed metallicity} are small. 
The $Y_{105}$-dropouts {($right\ panel$)} 
are undetected at $[3.6]$. The fit relies on the far-UV resulting  
in uncertain ages and masses. The gray dotted line shows the maximum 
$M/L$ allowed by the fit (95\% confidence). Clearly, moderately deeper 
IRAC would put stronger constraints on the allowed models at $z=8$. 
Upper limits are $2\sigma$. Note, the models were fitted to the fluxes
{of the $z\sim7$ and $z\sim8$  galaxies},  not the upper limits.
\label{figsed}}
\end{figure*}
}

\def\figmass{
\begin{figure}
\epsscale{1.23}
\centering
\plotone{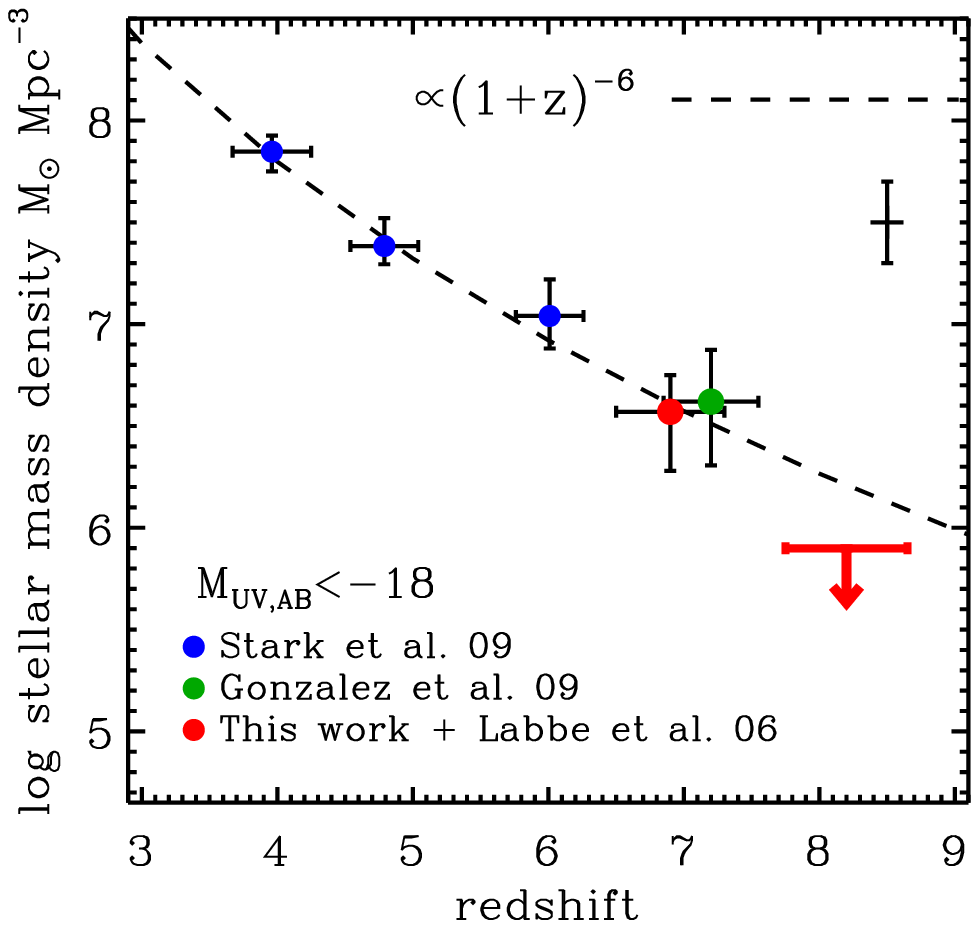}
\caption{Evolution of the integrated stellar mass density. 
The red circle shows the $z\sim7$ mass density, derived by
multiplying the integrated UV-luminosity density of \citet{Bo09a} 
with the mean $M/L$ of the sample studied here.
The other $z=3-7$ results are based on luminous samples ($M_{UV,AB}<-20$) 
from the literature (e.g., \citealt{St09}, $blue \
circles$; \citet{Go09}, $green \ circle$), corrected by +0.75 dex to 
$M_{UV,AB}<-18$, as appropriate for the $UV$ LF of \citet{Bo08} and a 
constant $M/L$.  The dashed line shows a $\propto(1+z)^{-6}$ evolution. 
The $z=8$ upper limit shows the 95\% confidence interval on the $M/L$, 
which is marginally in agreement with the extrapolated evolution. 
The floating error bar indicates the expected 
cosmic variance for the $z\sim7$ and $z\sim8$ samples. \label{figmass}}
\end{figure}
}

\begin{document}

\title{Ultradeep INFRARED ARRAY CAMERA  Observations Of Sub-$L^*$ $z\sim7$ And $z\sim8$ Galaxies in the
HUBBLE ULTRA DEEP FIELD: the Contribution of Low-Luminosity Galaxies to the Stellar
Mass Density and Reionization\altaffilmark{1}}

\author{I. Labb\'e\altaffilmark{2,3}, 
V. Gonz\'alez\altaffilmark{4}, 
R. J. Bouwens\altaffilmark{4,5}, 
G. D. Illingworth\altaffilmark{4}, 
P. A. Oesch\altaffilmark{6},
P. G. van Dokkum\altaffilmark{7}
C. M. Carollo\altaffilmark{6}, 
M. Franx\altaffilmark{5}, 
M. Stiavelli\altaffilmark{8},
M. Trenti\altaffilmark{9}, 
D. Magee\altaffilmark{4}, 
M. Kriek\altaffilmark{10}
}

\altaffiltext{1}{Based on observations made with the NASA/ESA {\it Hubble Space
Telescope}, which is operated by the Association of Universities for
Research in Astronomy, Inc., under NASA contract NAS 5-26555. These
observations are associated with programs \#11563, 9797. Based on observations 
with the {\em Spitzer Space Telescope}, which is operated by the Jet 
Propulsion Laboratory, California Institute of Technology under 
NASA contract 1407.  Support for this work was provided by NASA through contract 
125790 issued by JPL/Caltech. Based on service mode observations collected at 
the European Southern Observatory, Paranal, Chile (ESO Program 073.A-0764A).
Based on data gathered with the 6.5 m {\it Magellan} Telescopes located at Las Campanas 
Observatory, Chile.}
\altaffiltext{2}{Carnegie Observatories, Pasadena, CA 91101, USA}
\email{ivo@obs.carnegiescience.edu}
\altaffiltext{3}{Hubble Fellow}
\altaffiltext{4}{UCO/Lick Observatory, University of California, Santa Cruz, CA 95064, USA}
\altaffiltext{5}{Leiden Observatory, Leiden University, NL-2300 RA Leiden, The Netherlands}
\altaffiltext{6}{Institute for Astronomy, ETH Zurich, 8092 Zurich, Switzerland; poesch@phys.ethz.ch}
\altaffiltext{7}{Department of Astronomy, Yale University, New Haven, CT 06520, USA}
\altaffiltext{8}{Space Telescope Science Institute, Baltimore, MD 21218, USA}
\altaffiltext{9}{University of Colorado, Center for Astrophysics and Space Astronomy, 389-UCB, Boulder, CO 80309, USA}
\altaffiltext{10}{Department of Astrophysical Sciences, Princeton University,Princeton, NJ 08544, USA}

\begin{abstract}
We study the Spitzer Infrared Array Camera (IRAC) mid-infrared (rest-frame optical) fluxes of 14 newly 
WFC3/IR-detected $z\sim7$ $z_{850}-$dropout galaxies and 5 $z\sim8$ $Y_{105}$-dropout 
galaxies. The WFC3/IR depth and spatial resolution allow accurate removal of contaminating 
foreground light, enabling reliable flux measurements at $3.6\mu$m and $4.5\mu$m. 
None of the galaxies are detected to $[3.6]\approx26.9$ (AB,~$2\sigma$),
but a stacking analysis reveals a robust detection for the $z_{850}$-dropouts
and an upper limit for the $Y_{105}-$dropouts. 
We construct average broadband spectral energy distributions using the stacked ACS, WFC3, and IRAC fluxes and 
fit stellar  population synthesis models to derive mean redshifts, stellar masses, and ages.
For the $z_{850}-$dropouts, we find $z=6.9^{+0.1}_{-0.1}$, $(U-V)_{rest}\approx0.4$, 
reddening $A_V=0$, stellar mass $<M^*>=1.2^{+0.3}_{-0.6}\times10^9~M_\sun$ (Salpeter initial mass function). 
The best-fit ages $\sim300~$Myr, $M/L_V\approx0.2$, and $SSFR\sim1.7$~Gyr$^{-1}$ are 
similar to values reported for luminous $z\sim7$ galaxies, indicating the
galaxies are smaller but not much younger. The sub$-L^*$ galaxies observed here contribute 
significantly to the stellar mass density and under favorable conditions may have
provided enough photons for sustained reionization at $7<z<11$. In contrast, the 
$z=8.3^{+0.1}_{-0.2}$  $Y_{105}$-dropouts have stellar masses that are uncertain by 1.5 dex due to 
the near-complete reliance on far-UV data.  Adopting the $2\sigma$ upper limit on the 
$M/L(z=8)$, the stellar mass density to $M_{UV,AB}<-18$ declines from
$\rho^*(z=7)=3.7^{+1.4}_{-1.8}\times10^6~M_\sun~$Mpc$^{-3}$ to 
$\rho^*(z=8)<8\times10^5~M_\sun~$Mpc$^{-3}$, following $\propto(1+z)^{-6}$ 
over $3<z<8$. Lower masses at $z=8$ would signify more dramatic evolution, 
which can be established with deeper IRAC observations, long before 
the arrival of the James Webb Space Telescope.
\end{abstract}

\section{Introduction}
The reionization epoch represents the latest observational frontier of 
galaxy formation. Little is known about the galaxy properties during
this time, including feedback, metal production, and their 
contribution to reionization. Extensive studies have been done of galaxies 
selected by the dropout technique out to $z=6$ (see e.g., \citealt{St03,
Ya04, Bo06, Mc09a}). Pushing these studies to $z\gtrsim7$ has 
proven extraordinarily challenging: only $\sim$25 high-quality 
candidates are currently known \citep[][R. J. Bouwens et al. in preparation]{Bo08,Oe09a,Ca09,Ou09}.

The arrival of WFC3/IR aboard {\it Hubble Space Telescope (HST)}
heralds a dramatic improvement in our ability to survey the 
reionization era, allowing us to identify the predominant
low-luminosity galaxies at $z\gtrsim7$ by their redshifted $UV$ light.
 Whereas {\it HST} remains crucial for identifying the galaxies, longer-wavelength data are 
necessary for constraining the stellar 
masses and ages. Access to the rest-frame optical is offered by the Infrared 
Array Camera (IRAC; \citealt{Fa04}) on {\it Spitzer}, which by itself 
has permitted estimates of stellar masses for large numbers of $z\sim5-6$ sources 
\citep{Ey05, Ya06, St09}, and even a few at $z\sim7$ \citep{Eg05,La06}. One 
surprising early finding was the number of quite massive $\sim10^{10}M_\sun$
galaxies with appreciable ages ($200-300$Myr) suggesting the galaxies 
formed substantial amounts of stars at even earlier times, well into 
the epoch of reionization \citep{St07, Ya06}.

\figstack

In this Letter, we extend the stellar mass and age estimates to lower 
luminosities ($0.06~L^*_{z=3}$) and higher redshift $z\approx8$ by analyzing the 
IRAC mid-IR fluxes of the robust $z_{850}$-dropout and $Y_{105}$-dropout galaxy samples
found by \citet{Oe09b} and \citet{Bo09a} in the newly acquired WFC3/IR data  (GO-11563, PI Illingworth) 
over the Hubble Ultra Deep Field (HUDF; see also \citealt{Bu09,Mc09b}). 
We adopt an $\Omega_M=0.3, \Omega_\Lambda=0.7,$ cosmology with 
$H_0=70$~km~s$^{-1}$Mpc$^{-1}$. Magnitudes are in the AB photometric system \citep{Ok83}.
Imaging depths refer to $1\sigma$ AB point source total magnitude, unless
stated otherwise.

\section{Observations and Photometry}
The recent $z=7$ \zdrop \  and $z=8$ \ydrop \ 
samples were selected from exceptionally deep WFC3/IR imaging,
to $\approx29$ AB magnitude $5\sigma$ in \yy,\jj,\hh,  allowing the
study of fainter sources than the NICMOS based samples  presented in 
\citet{La06} and \citet{Go09}. We discuss {briefly} the IRAC photometry 
of 14 of the 16 $z_{850}-$dropouts found by \cite{Oe09b} and the 5 $Y_{105}-$dropouts 
reported by \citet{Bo09a} (see \citealt{La06} for the two $z_{850}-$dropouts found 
earlier with NICMOS).

{\it Spitzer} data over this field were obtained from the
Great Observatories Origins Deep Survey (GOODS; M. Dickinson et al. in preparation),
which observed the HUDF with IRAC in two epochs of $\approx23.3$ hr each\footnote{This 
paper uses data release DR3 of epoch 1 and data release DR2 of epoch 2, 
available from \url{http://data.spitzer.caltech.edu/popular/goods/}}. 
The IRAC depths in the $3.6$ and $4.5\micron$ bands are 27.7 and 27.2 (AB). 
We supplement the observations with deep $K_s$-band
data from the {\it Very Large Telescope} and {\it Magellan} to 
K$_{s,AB}=27.4$ (I. Labb\'e et al. in preparation).

A critical aspect is obtaining reliable IRAC fluxes
of the candidates. This is challenging because of extended wings of the 
IRAC point-spread function (PSF), causing flux contamination from nearby foreground 
sources. We use the technique of \citet{La06}  to fit and
subtract the foreground sources  using the flux profiles
in the ultradeep WFC3/IR detection images as priors (see 
\citealt{Go09} for a more complete description 
and \citealt{Wu07} for illustrative examples).

After cleaning the IRAC images, we first performed conventional
aperture photometry in $3.6$ and $4.5\micron$ bands
in 2\farcs5 diameter apertures. 
We define the photometric error as the quadratic sum of random fluctuations in the aperture 
(determined from empty apertures on the nearby background; e.g., 
\citealt{La03}) and the fit uncertainties to each individual neighbor.
We performed the photometry independently on the first and second epoch 
IRAC data, which are rotated by $180\deg$ and therefore have different
PSFs and contamination (see \citealt{La06}).
Two of the 14 $z_{850}$-dropouts and 1 $Y_{105}$-dropout
were too close to bright sources to be successfully measured and were removed
from the sample (later we assume average $M/L$ for these galaxies).
Of the remaining galaxies none were detected ($>2\sigma$) 
in the epochs individually or combined (to $[3.6]>26.9$, $2\sigma$ total). 

\figsed

\section{Stacking}
Owing to the extreme faintness of the \zdrop \ and \ydrop \ samples
over the WFC3 HUDF09 field, we derive average 
properties by registering and stacking the confusion corrected
IRAC maps. The resulting stacked image for the 12 $z_{850}$-dropouts is shown in Figure~\ref{figstack},
featuring an unambiguous source coincident with the \zdrop \ location,
with an average magnitude of $[3.6]=27.2$ (AB, measured in a 2\farcs5 
diameter aperture and corrected to total).
To evaluate robustness of the detection we also create a median 
stack, which is resistant against outliers, to ensure that the 
signal is not coming from only a few sources. We derive 
uncertainties by bootstrap resampling the images in the stack 200 times, 
finding  signal-to-noise ratio $S/N\approx5.1$. The implied $1\sigma$ rms 
in a 2\farcs5 aperture is $[3.6]=29.0$ (total). The same procedure was 
carried out in the $[4.5]-$band showing a fainter source ($SNR=2.6$).

Detecting the \ydrop s is even more challenging. Not only are the
$Y_{105}$-dropouts in our samples fainter, there are also fewer to 
stack (4 versus 12), and they might be bluer due to younger ages, 
lower metallicities, or other effects (e.g., \citealt{Bo09b}). 
The resulting  stacked image for the 4 is shown in the lower panels of 
Figure~\ref{figstack}, showing no detection to a limit $[3.6]>27.5$ ($2\sigma$, total). The formal
measurement is $S/N=0.9$.  The photometry is presented in Table~1.

The average SED shape of the faint $z\sim7$ \zdrop \ galaxies, shown in 
Figure~\ref{figsed}, is remarkably similar
to that of the more luminous and massive NICMOS sample presented by 
\cite{Go09}. The rest-frame far-UV slope is extremely blue 
(\citealt{Bo09b} find $\beta=-3.0\pm0.2$), suggesting low ages or sub-Solar metallicities. 
However, note the relatively red \hh$-[3.6]=0.7^{+0.2}_{-0.25}$ color, suggesting
the presence of a modest Balmer break between the bands, indicative of evolved stellar populations.

\section{Stellar populations}

We use standard techniques to derive stellar masses and redshifts 
by fitting stellar populations synthesis models to the average 
$i,z,Y,J,H,K,3.6,4.5$-band flux densities of the \zdrop s and \ydrop s.
We use \citet[][BC03]{bc03} models and assume Solar metallicity
and a \citet{Sa55} initial mass function (IMF) 
between $0.1$ and $100$~\msun. This choice enables  
straightforward comparison with previous results, but we also
review the effect of different assumptions and models. Furthermore, we 
assume constant star formation (CSF) as opposed to exponentially declining
models, motivated by the lack of evolution of the star formation rate (SFR) per unit mass
(specific SFR,(SSFR)) between $z=3$ and $7$ \citep{Go09,St09}. Up to $A_V=0.5$~mag 
\citet{Cal00} starburst reddening is allowed, where we note that $z\gtrsim5$ galaxies 
have blue UV-continuum slopes and low dust extinction 
(e.g., \citealt{Le03, St05, Bo06,Bo09b}). We use 
the $\chi^2$-fitting code FAST \citep{Kr09} which provides best-fit parameters 
and uncertainties using Monte Carlo simulations. We fit the average 
SED fluxes to models smoothed to a resolution of 100~\AA\ rest frame, 
corresponding to the approximate width of the dropout selection windows.
The modeling results are presented in Table~2 and overplotted in Figure~\ref{figsed}. 

The best-fit redshifts of the $z_{850}-$dropouts and $Y_{105}-$dropouts are $z=6.9^{+0.1}_{0.1}$ and 
$z=8.3^{+0.1}_{-0.2}$, respectively, in agreement with expectations from simulations 
(see \citealt{Oe09b,Bo09a}).  The extremely blue rest-frame far-UV slope of the $z\sim7$
galaxies is a challenge to fit, requiring $A_V=0$, and sub-Solar metallicity rather than Solar models
($\chi^2_{red}=2.9$ versus $\chi^2_{red}=6.4$). Consistently, the relatively red
$(U-V)_{rest}\approx0.4$ color favors evolved models 
(age$_w=350^{+30}_{-170}~$Myr\footnote{Following \cite{La06} and \cite{Go09}, 
we report {\em SFH weighted} age$_{w}=t/2$ (for CSF), where $t$ is the time 
elapsed since the start of star formation.}). 
We caution that emission lines can contribute to the $[3.6]$ flux, 
but for the redshift distribution and  $H_{160}-[3.6]$ color of this sample we estimate
the effect is only {$\sim0.05-0.1$ mag}.

Interestingly, \citet{Ma05} and Charlot and Bruzual (in preparation; CB07) 
models produce nearly identical results as the BC03 models. Derived stellar masses are 
quite robust with at most 0.1~dex systematic variation between the different 
model choices. The only important effect on the masses is 
IMF: a more reasonable \citet{Kr01} IMF reduces the
stellar masses and SFRs by $0.2$~dex, but does 
not change other parameters or the quality of fit. More detailed exploration 
of the stellar ages and star formation histories (SFHs) is presented by \cite{Go09}.

The stellar masses and ages of the \ydrop \ galaxies are not well constrained. 
The low S/N of the IRAC stack causes the fit to be determined by
the \jj \ and \hh\ fluxes, producing a formal best-fit with minimal allowed masses 
$\sim4\times10^{7}M_\sun$ and ages $\sim10$~Myr and $\sim$1.5 dex uncertainties. 
However, if we refit the $z\sim7$ stack without IRAC data we also find 
very low masses and ages and large uncertainties. This underscores the 
crucial importance of deep IRAC data and suggests that model fits relying 
exclusively on far-UV data should be regarded with caution. 
Even though the fits may suggest a strong evolution of the 
stellar populations, we can not exclude that galaxies at $z\sim8$ have 
similar masses, ages, and $M/L$s as faint  $z\sim7$ galaxies.

\figmass

\section{Discussion}
The reionization epoch represents the latest frontier of galaxy formation theory, 
which is now being probed to extreme depths with the amazingly efficient WFC3/IR 
camera aboard {\it HST}. Using the conservative $z_{850}-$dropout and \ydrop \ samples of 
\citet{Oe09b} and \citet{Bo09a} we find {\it Spitzer}/IRAC capable of an equally 
remarkable feat: the stacked $3.6\micron$ detection of even the faintest 
$z\approx6.9$ galaxies found by WFC3/IR, providing direct proof that IRAC 
can probe much deeper than widely accepted. In addition we derive upper limits for the 
newly discovered $z\approx8.3$ galaxies, some of which may well be at higher redshift than 
the currently most distant $z\approx8.2\pm0.1$ GRB \citep{Sa09,Ta09}. We 
now elaborate on various implications.\\

{\it 1) Stellar populations at $z\sim7$} \\
{\it Spitzer}/IRAC places two important constraints on the stellar populations. First, the
combination of the blue far-UV slope and the moderately red $H_{160}-[3.6]=0.7^{+0.2}_{-0.25}$ color 
in the ultrafaint ($H_{160}\approx27.9$)  $z\sim7$ galaxies indicates the presence of 
a modest Balmer break, expected for evolved  stellar populations. Very young 
stellar ages are ruled out (CSF age$_w\gtrsim80~$Myr at $95$\%).
Interestingly, low-metallicity models, which produce bluer far-UV colors at fixed 
age, provide significantly better fits than do Solar, but for both it is 
challenging to reproduce the extremely blue observed far-UV slope ($\beta=-3.0\pm0.2$ as found 
by \citealt{Bo09b}).
Second, the average stellar masses $<M> \sim 1\times10^{9}~M_\sun$ indicate appreciable 
mass-to-light ratios M/L$_{1500}\approx0.1$ and M/L$_V\approx0.2$. The ratios 
are comparable those of the more massive $\sim7\times 10^9~M_\sun$ sample of \cite{Go09}, 
suggesting that the $M/L$ ratios do not strongly depend on luminosity or mass. 
We find no appreciable difference between the best-fit parameters of BC03, MA05, and 
CB07 models, although IMF variations are a potentially significant source of systematic 
uncertainty: a steeper high-mass slope (e.g., \citealt{We05}) would lower the 
inferred stellar ages, a higher characteristic mass (e.g., ``bottom light'', \citealt{vD08}) 
would lower the total stellar masses and $M/L$s, and any 
evolution with time would complicate comparisons between SFR and stellar mass 
across epochs.\\

{\it 2) Star formation histories} \\
SFHs are notoriously hard to constrain from observations of individual
galaxies. However, statistical constraints can be obtained by comparing the SSFRs 
as a function of time and mass, as different shapes of the
SFH give rise to distinctly different evolution (e.g., \citealt{La07}). One 
remarkable recent result is the observation that the SSFR at fixed $\sim5\times10^9 M_\sun$ 
does not evolve from $z=3-7$ \citep{St09,Go09}. Here we benefit from the extreme 
depths of the WFC3/IR data to probe the lowest luminosities, suggesting that 
the SSFR at $z\sim7$ is also not strongly
dependent on luminosity or stellar mass over the range $1-7\times10^9 M_\sun$. This may imply
that the SFR correlates with stellar mass with a logarithmic slope close to 1. 
Both results (with redshift and mass) qualitatively agree with numerical 
simulations for early galactic star formation ($z>4)$, which
robustly predict that SFR is proportional to M*, with similar SFHs in haloes of different
masses \citep{Da08}, and that the SFRs of individual galaxies are constantly 
rising \citep{Fi07}.\\

{\it 3) Evolution of the stellar mass density to $z=7-8$} \\
The low-luminosity galaxies probed in this paper are expected to contribute significantly
to the stellar mass density owing to their substantial $M/L$. 
Following the approach of \citet{Go09}, we derive integrated stellar mass 
densities at $z=7-8$ by multiplying the UV-luminosity densities of \citet{Bo09a}, integrated 
to $M_{UV,AB}=-18$, with the mean $M/L$ derived for the galaxy samples in this Letter.
For the \zdrop s at $z\sim7$, this results in $\rho^*=3.7^{+1.4}_{-1.8}\times10^6~M_\sun $Mpc$^{-3}$.
The evolution of the mass density is shown in Figure~\ref{figmass}, compared to previous 
luminous samples, which were corrected to the same luminosity limit assuming a constant $M/L$.
The stellar masses at $z=8$ are uncertain by $\sim1.5~$dex due to the weak constraints
from IRAC, so instead of plotting the best-fit $M/L$ we show the  95\% confidence upper limit.
The integrated mass density  to a limit of $M_{UV,AB}<-18$
then declines to $<8\times10^5~M_\sun$Mpc$^{-3}$. This is 
marginally consistent with the extrapolation of the $\propto(1+z)^{-6}$ evolution over the 
range  $3<z<7$. The upper limit may be evidence for a more rapid evolution 
toward $z=8$, but deeper IRAC data and larger samples are needed to establish this firmly.
The uncertainties quoted above do not include cosmic variance, which is estimated to be
$\sim30$\% for the $z\sim7$ sample and $\sim40$\% at $z\sim8$ \citep{Tr08}.\\

{\it 4) Reinionization since $z\sim11$} \\
Combined, the results have profound implications for the ability
of the universe to be reionized by photons from star formation -- 
and to remain so for an extended period. Crucially, the 
large stellar mass density in $z\sim7$ low-luminosity galaxies points to 
significant amounts of star formation at earlier times.
We can therefore ask whether this corresponds to enough
ionizing photons at higher redshift to keep the universe reionized. 
Let us assume the stellar mass density $\rho^*=3.7\times10^6~M_\sun$Mpc$^{-3}$ was assembled 
in the $340$~Myr between $7<z<11$, in the reionization era (WMAP5, \citealt{Ko09}). 
Correcting by 10\% to account for mass loss in stellar evolution, the possible sustained SFR density
would then be $\rho_{SFR}\approx0.012~M_\sun$yr$^{-1}$Mpc$^{-3}$, significantly 
higher than that has been derived for luminous galaxies alone \citep{Go09}. 
However, the primary uncertainty is the fiducial critical SFR
density needed for reionization $\rho^{SFR}_{crit}\propto~C/f_{esc}$ \citep{Ma99}, 
which depends sensitively on the values for the \ion{H}{2} IGM clumping factor $1<C<30$ and
the fraction of ionizing photons $0.05<f_{esc}<1$ that leak unhindered out of 
the galaxies (e.g., \citealt{Ou09}). The popular choice $(C=30,f_{esc}=0.1)$ leads to 
$<\rho^{SFR}_{crit}>=0.7~M_\sun$yr$^{-1}$Mpc$^{-3}$ averaged over $7<z<11$, 
which is $60$ times higher than that 
can be explained by the stellar masses. However, at early times the clumping factor may
be as low as $C=3-6$ (e.g., \citealt{Tr07,Bo07,Pa09}). The escape fraction
may be significantly higher as well: recent numerical simulations suggest 
$f_{esc}\gtrsim0.2$ \citep{Wi09,Pa09,Ya09} and \citet{Bo09b} note that the extremely 
blue far-UV slope of the galaxies studied here could also result from large escape 
fractions $f_{esc}\gtrsim0.3$. The choice $(C=3,f_{esc}=0.5)$ leads to 
$<\rho^{SFR}_{crit}>\approx0.014~M_\sun$yr$^{-1}$Mpc$^{-3}$(over $7<z<11$), 
which means photons from the low-luminosity star forming galaxies 
observed here are capable of causing sustained reionization since $z=11$. \\

The first {\it Spitzer}/IRAC detection of low-luminosity $z=7$ galaxies 
and their red $H-[3.6]\approx0.7$ colors
provide important insights into the earliest phases of galaxy evolution, showing that
these early galaxies have relatively high $\sim300~$Myr ages and $M/L_V\approx0.2$. 
Consequently, low-luminosity galaxies contribute significantly to 
the stellar mass density $\rho^*(z=7)=3.7^{+1.4}_{-1.8}\times10^6~M_\sun $Mpc$^{-3}$,
suggesting that the SFR corresponding to that mass may have provided a substantial
fraction of the ionizing photons that kept the universe reionized at earlier times.
Future IRAC studies of larger numbers of high redshift galaxies are required 
to bolster the results in this paper and to further constrain the evolution of the mass 
density toward the highest redshifts $z\gtrsim8$. The {\it Spitzer} post-cryogenic phase 
(``warm mission'') provides a rare window of opportunity to observe these galaxies in a timely 
manner to the $\sim0.5~$mag deeper limits needed, long before the arrival of the
{\it James Webb Space Telescope}.\\

\acknowledgments
 
We are grateful to all those at NASA, STScI, JPL, SSC, and throughout the community
who have worked so diligently to make Hubble and Spitzer the remarkable 
observatories that they are today. 
We thank Kristian Finlator, Masami Ouchi, and Risa Wechsler for stimulating discussion.
IL acknowledges support from NASA through Hubble Fellowship grant 
HF-01209.01-A awarded by the STScI, which is operated by the AURA, Inc., 
for NASA, under contract NAS 5-26555. PO acknowledges support from the Swiss 
National Foundation (SNF). We acknowledge the support of NASA grant NAG5-7697 and NASA grant 
HST-GO-11563.01. \\

{\it Facilities:} \facility{{\it Spitzer} (IRAC)}, \facility{{\it HST} (WFC3/IR)}, \facility{{\it Magellan:Baade} (PANIC)}, \facility{{\it VLT:Melipal} (ISAAC)}

\begin{deluxetable}{crrrrrrrrrr}
\tabletypesize{\scriptsize}
\tablecaption{Stacked photometry of $z\sim7$ and $z\sim8$ dropout galaxies in the HUDF}
\tablewidth{0pt}
\tablehead{
\colhead{} & \colhead{$B_{435}$}  & \colhead{$V_{606}$}  & \colhead{$i_{775}$} 
& \colhead{$z_{850}$} & \colhead{$Y_{105}$} & \colhead{$J_{125}$} & 
\colhead{$H_{160}$}  & \colhead{$K_{s}$}& \colhead{[3.6]} & \colhead{[4.5]} }
\startdata
$z_{850}$-dropout SED & 1.4(0.7)&0.4(0.4)&-0.1(0.6)&5.6(1.1)&27.2(1.8)
&29.6(2.0)&24.5(1.6)&22.9(13.8)&48.5(9.5)&43.2(16.6) \\
\tableline \\ 
$Y_{105}$-dropout SED & -1.1(0.9)&0.2(0.7)&0.3(0.8)&-0.8(1.4)&5.0(1.4)
&20.9(1.4)&18.8(1.3)&15.1(20.7)&16.9(18.3)&20.5(25.0) 
\enddata
\tablecomments{The optical--to--near-IR photometry is measured in $0\farcs4$ diameter 
apertures and corrected to total. {\it Spitzer}/IRAC photometry is performed on confusion corrected maps in
$2\farcs5$ diameter apertures and corrected to total. Note, we limit the signal-to-noise ratio to
$S/N\approx15$ or 0.07 mag to account for zeropoint uncertainties. Units are nanoJy.}
\end{deluxetable}

\begin{deluxetable}{ccccccccc}
\tabletypesize{\scriptsize}
\tablecaption{Best-fit stellar population parameters \\ for constant star forming models and a Salpeter (1955) IMF}
\tablewidth{0pt}
\tablehead{
\colhead{model} & \colhead{z$_{phot}$} & \colhead{Z} & \colhead{log age$_w$} & \colhead{A$_V$} &
 \colhead{log $M^*$} & \colhead{log $SFR$} & \colhead{log $SSFR$}  & \colhead{$\chi^2_{red}$}\\
\colhead{} & \colhead{} & \colhead{(Z$_\sun$)} & \colhead{(yr)} & \colhead{(mag)} &
 \colhead{($M_\sun$)} & \colhead{($M_\sun~$yr$^{-1}$)} & \colhead{(yr$^{-1}$)}  & \colhead{}
}
\startdata
 \multicolumn{8}{c}{ $z_{850}$-Dropouts Solar Metallicity } \\ 
 \tableline \\ 
BC03 & 6.85$^{+0.08}_{-0.07}$  & 1.00 & 8.58$^{+0.00}_{-0.28}$  & 0.00$^{+0.08}_{-0.00}$  & 9.15$^{+0.07}_{-0.28}$  & 0.37$^{+0.07}_{-0.02}$  & -8.78$^{+0.21}_{-0.00}$  & 6.39 \\ 
CB07 & 6.85$^{+0.07}_{-0.07}$  & 1.00 & 8.58$^{+0.00}_{-0.30}$  & 0.00$^{+0.05}_{-0.00}$  & 9.15$^{+0.05}_{-0.30}$  & 0.37$^{+0.05}_{-0.02}$  & -8.78$^{+0.24}_{-0.00}$  & 6.39 \\ 
MA05 & 6.87$^{+0.07}_{-0.07}$  & 1.00 & 8.45$^{+0.13}_{-0.36}$  & 0.00$^{+0.08}_{-0.00}$  & 9.10$^{+0.16}_{-0.36}$  & 0.43$^{+0.08}_{-0.02}$  & -8.67$^{+0.32}_{-0.12}$  & 5.48 \\ 
 \tableline \\ 
 \multicolumn{8}{c}{ $z_{850}$-Dropouts sub-Solar Metallicity } \\ 
 \tableline \\ 
BC03 & 6.88$^{+0.07}_{-0.07}$  & 0.20 & 8.58$^{+0.00}_{-0.32}$  & 0.00$^{+0.10}_{-0.00}$  & 9.09$^{+0.07}_{-0.29}$  & 0.31$^{+0.12}_{-0.01}$  & -8.78$^{+0.25}_{-0.00}$  & 2.95 \\ 
CB07 & 6.88$^{+0.08}_{-0.07}$  & 0.20 & 8.55$^{+0.03}_{-0.30}$  & 0.00$^{+0.12}_{-0.00}$  & 9.07$^{+0.10}_{-0.29}$  & 0.31$^{+0.14}_{-0.01}$  & -8.76$^{+0.23}_{-0.02}$  & 2.94 \\ 
MA05 & 6.88$^{+0.08}_{-0.07}$  & 0.05 & 8.55$^{+0.03}_{-0.32}$  & 0.00$^{+0.12}_{-0.00}$  & 9.07$^{+0.10}_{-0.28}$  & 0.31$^{+0.15}_{-0.01}$  & -8.76$^{+0.23}_{-0.02}$  & 2.72 \\ 
 \tableline \\ 
 \multicolumn{8}{c}{ $Y_{105}$-Dropouts Solar Metallicity } \\ 
 \tableline \\ 
BC03 & 8.21$^{+0.13}_{-0.19}$  & 1.00 & 6.70$^{+1.77}_{-0.00}$  & 0.00$^{+0.38}_{-0.00}$  & 7.57$^{+1.47}_{-0.03}$  & 0.58$^{+0.36}_{-0.27}$  & -6.99$^{+0.00}_{-1.70}$  & 2.31 \\ 
CB07 & 8.21$^{+0.14}_{-0.20}$  & 1.00 & 6.70$^{+1.77}_{-0.00}$  & 0.00$^{+0.40}_{-0.00}$  & 7.57$^{+1.51}_{-0.03}$  & 0.58$^{+0.39}_{-0.27}$  & -6.99$^{+0.00}_{-1.70}$  & 2.31 \\ 
MA05 & 8.23$^{+0.15}_{-0.23}$  & 1.00 & 6.70$^{+1.80}_{-0.00}$  & 0.00$^{+0.45}_{-0.00}$  & 7.57$^{+1.55}_{-0.03}$  & 0.59$^{+0.44}_{-0.23}$  & -6.98$^{+0.00}_{-1.74}$  & 2.20 \\ 
 \tableline \\ 
 \multicolumn{8}{c}{ $Y_{105}$-Dropouts sub-Solar Metallicity } \\ 
 \tableline \\ 
BC03 & 8.26$^{+0.15}_{-0.26}$  & 0.20 & 6.70$^{+1.80}_{-0.00}$  & 0.00$^{+0.50}_{-0.00}$  & 7.59$^{+1.45}_{-0.02}$  & 0.60$^{+0.50}_{-0.36}$  & -6.99$^{+0.00}_{-1.72}$  & 1.57 \\ 
CB07 & 8.26$^{+0.14}_{-0.25}$  & 0.20 & 6.70$^{+1.77}_{-0.00}$  & 0.00$^{+0.50}_{-0.00}$  & 7.59$^{+1.43}_{-0.02}$  & 0.60$^{+0.49}_{-0.36}$  & -6.99$^{+0.00}_{-1.69}$  & 1.57 \\ 
MA05 & 8.26$^{+0.14}_{-0.25}$  & 0.05 & 6.70$^{+1.77}_{-0.00}$  & 0.00$^{+0.50}_{-0.00}$  & 7.58$^{+1.44}_{-0.02}$  & 0.60$^{+0.49}_{-0.36}$  & -6.99$^{+0.00}_{-1.70}$  & 1.62 
\enddata
\tablecomments{
Best-fit parameters and 68\% confidence intervals as computed with 
FAST \citep{Kr09}  for CSF models with a Salpeter (1955) IMF between $0.1-100~M_\sun$.
A Kroupa (2001) IMF would result in 0.20~dex lower stellar
masses and SFRs, respectively, but other parameters would remain unchanged. The 
\zdrop\ model fits favor low metallicities and high ages. The \ydrop \ model fits are 
uncertain: deeper IRAC is needed. Note, following \cite{Go09} we report the SFH 
averaged $age_w=t/2$ (for CSF), where $t$ is the time elapsed since the 
start of star formation}
\end{deluxetable}


\begin{thebibliography}{}
\bibitem[Bolton \& Haehnelt(2007)]{Bo07} Bolton, J.~S., \& Haehnelt, M.~G.\ 2007, \mnras, 374, 493 
\bibitem[Bouwens et al.(2006)]{Bo06} Bouwens, R.~J., Illingworth, G.~D., Blakeslee, J.~P., \& Franx, M.\ 2006, \apj, 653, 53 
\bibitem[Bouwens et al.(2008)]{Bo08} Bouwens, R.~J., Illingworth, G.~D., Franx, M., \& Ford, H.\ 2008, \apj, 686, 230 
\bibitem[Bouwens et al.(2009a)]{Bo09a} Bouwens, R.~J., et al.\ 2009a, arXiv:0909.1803
\bibitem[Bouwens et al.(2009b)]{Bo09b} Bouwens, R.~J., et al.\ 2009b, ApJ, in press, arXiv:0910.0001 
\bibitem[Bunker et al.(2009)]{Bu09} Bunker, A., et al.\ 2009, arXiv:0909.2255 
\bibitem[Bruzual \& Charlot(2003)]{bc03}  Bruzual, G.~\& Charlot, S.\ 2003, \mnras, 344, 1000 (BC03)
\bibitem[Calzetti et al.(2000)]{Cal00} Calzetti, D., et al.\ 2000, \apj, 533, 682 
\bibitem[Castellano et al.(2009)]{Ca09} Castellano, M., et al.\ 2009, arXiv:0909.2853 
\bibitem[Dav{\'e}(2008)]{Da08} Dav{\'e}, R.\ 2008, \mnras, 385, 147 
\bibitem[Eyles et al.(2005)]{Ey05} Eyles, L.~P., et al.\ 2005, \mnras, 364, 443 
\bibitem[Egami et al.(2005)]{Eg05} Egami, E., et al.\ 2005, \apjl, 618, L5 
\bibitem[Fazio et al.(2004)]{Fa04} Fazio, G.~G., et al.\  2004, \apjs, 154, 10 
\bibitem[Finlator et al.(2007)]{Fi07} Finlator, K., Dav{\'e}, R., \& Oppenheimer, B.~D.\ 2007, \mnras, 376, 1861 
\bibitem[Gonzalez et al.(2009)]{Go09} Gonzalez, V., Labbe, I., Bouwens, R.~J., Illingworth, G., Franx, M., Kriek, M., \& Brammer, G.~B.\ 2009, arXiv:0909.3517
\bibitem[Kriek et al.(2009)]{Kr09} Kriek, M., van Dokkum, P.~G., Labb{\'e}, I., Franx, M., Illingworth, G.~D., Marchesini, D., \& Quadri, R.~F.\ 2009, \apj, 700, 221 
\bibitem[Kroupa(2001)]{Kr01} Kroupa, P.\ 2001, \mnras, 322, 231 
\bibitem[Komatsu et al.(2009)]{Ko09} Komatsu, E., et al.\ 2009, \apjs, 180, 330 
\bibitem[Labb{\'e} et al.(2003)]{La03} Labb{\'e}, I., et al.\ 2003, \aj, 125, 1107
\bibitem[Labb{\'e} et al.(2006)]{La06} Labb{\'e}, I., Bouwens, R., Illingworth, G.~D., \& Franx, M.\ 2006, \apjl, 649, L67 
\bibitem[Labb{\'e} et al.(2007)]{La07} Labb{\'e}, I., et al.\ 2007, \apj, 665, 94 
\bibitem[Lehnert \& Bremer(2003)]{Le03} Lehnert, M.~D., \& Bremer, M.\ 2003, \apj, 593, 630 
\bibitem[Madau et al.(1999)]{Ma99} Madau, P., Haardt, F., \& Rees, M.~J.\ 1999, \apj, 514, 648 
\bibitem[Maraston(2005)]{Ma05} Maraston, C.\ 2005, \mnras, 362, 799 
\bibitem[McLure et al.(2009a)]{Mc09a} McLure, R.~J., Cirasuolo, M., Dunlop, J.~S., Foucaud, S., \& Almaini, O.\ 2009, \mnras, 395, 2196 
\bibitem[McLure et al.(2009b)]{Mc09b} McLure, R.~J., Dunlop, J.~S., Cirasuolo, M., Koekemoer, A.~M., Sabbi, E., Stark, D.~P., Targett, T.~A., \& Ellis, R.~S.\ 2009b, MNRAS, in press, arXiv:0909.2437 
\bibitem[Oesch et al.(2009a)]{Oe09a} Oesch, P.~A., et al.\ 2009a, \apj, 690, 1350
\bibitem[Oesch et al.(2009b)]{Oe09b} Oesch, P.~A., et al.\ 2009b, ApJ, in press, arXiv:0909.1806 
\bibitem[Oke \& Gunn(1983)]{Ok83} Oke, J.~B., \& Gunn, J.~E.\ 1983, \apj, 266, 713 
\bibitem[Ouchi et al.(2009)]{Ou09} Ouchi, M., et al.\ 2009, arXiv:0908.3191 
\bibitem[Pawlik et al.(2009)]{Pa09} Pawlik, A.~H., Schaye, J., \& van Scherpenzeel, E.\ 2009, \mnras, 394, 1812 
\bibitem[Salpeter(1955)]{Sa55} Salpeter, E.~E.\ 1955, \apj, 121, 161 
\bibitem[Stanway et al.(2005)]{St05} Stanway, E.~R., McMahon, R.~G., \& Bunker, A.~J.\ 2005, \mnras, 359, 1184 
\bibitem[Salvaterra et al.(2009)]{Sa09} Salvaterra, R., et al.\ 2009, arXiv:0906.1578
\bibitem[Stanway et al.(2003)]{St03} Stanway, E.~R., Bunker, A.~J., \& McMahon, R.~G.\ 2003, \mnras, 342, 439 
\bibitem[Stark et al.(2009)]{St09} Stark, D.~P., Ellis, R.~S., Bunker, A., Bundy, K., Targett, T., Benson, A., \& Lacy, M.\ 2009, \apj, 697, 1493 
\bibitem[Stark et al.(2007)]{St07} Stark, D.~P., Ellis, R.~S., Richard, J., Kneib, J.-P., Smith, G.~P., \& Santos, M.~R.\ 2007, \apj, 663, 10 
\bibitem[Tanvir et al.(2009)]{Ta09} Tanvir, N.~R., et al.\ 2009, arXiv:0906.1577
\bibitem[Trenti \& Stiavelli(2008)]{Tr08} Trenti, M., \& Stiavelli, M.\ 2008, \apj, 676, 767 
\bibitem[Trac \& Cen(2007)]{Tr07} Trac, H., \& Cen, R.\ 2007, \apj, 671, 1 
\bibitem[Yajima et al.(2009)]{Ya09} Yajima, H., Umemura, M., Mori, M., \& Nakamoto, T.\ 2009, \mnras, 398, 715 
\bibitem[Yan et al.(2004)]{Ya04} Yan, H., et al.\ 2004, \apj, 616, 63 
\bibitem[Yan et al.(2006)]{Ya06} Yan, H., Dickinson, M., Giavalisco, M., Stern, D., Eisenhardt, P.~R.~M., \& Ferguson, H.~C.\ 2006, \apj, 651, 24 
\bibitem[van Dokkum(2008)]{vD08} van Dokkum, P.~G.\ 2008, \apj, 674, 29 
\bibitem[Weidner \& Kroupa(2005)]{We05} Weidner, C., \& Kroupa, P.\ 2005, \apj, 625, 754 
\bibitem[Wise \& Cen(2009)]{Wi09} Wise, J.~H., \& Cen, R.\ 2009, \apj, 693, 984
\bibitem[Wuyts et al.(2007)]{Wu07} Wuyts, S., et al.\ 2007, \apj, 655, 51 
\end{thebibliography}
\end{document}